\documentclass[11pt]{article}
\usepackage{moriond,graphicx}

\def\Journal#1#2#3#4{{#1} {\bf #2}, #3 (#4)}

\def\NPB{{\em Nucl. Phys.} B}
\def\PLB{{\em Phys. Lett.}  B}
\def\PRL{{\em Phys. Rev. Lett.}}

\def\EPJ{{\em Eur. Phys. J.} C}

\def\be{\begin{equation}}
\def\ee{\end{equation}}
\def\bea{\begin{eqnarray}}
\def\eea{\end{eqnarray}}

\begin{document}
\vspace*{4cm}
\title{THE LEP TESTIMONY: EXOTIC SEARCHES AND STUDIES}

\author{ GABRIELLA P\'ASZTOR 
\footnote{On leave of absence from 
KFKI RMKI, Budapest, Hungary.}
}

\address{Department of Physics, University of California, Riverside, CA 92521,
USA \\
Mailing address: CERN, Gen\`eve 23, CH-1211, Switzerland \\
E-mail: {\tt Gabriella.Pasztor{\protect @}cern.ch}}

\maketitle
\abstracts{ A selection of recent results on searches for phenomena
beyond the Standard Model 
is presented from the LEP Collaborations, based on the data
collected up to the highest centre-of-mass energies of 209 GeV.}

\section{Introduction}\label{sec:intro}

The Standard Model (SM) accurately describes the observed phenomena, but it
leaves several fundamental questions unanswered. Many extensions of the SM have
been developed to solve these puzzles, of which the supergravity inspired
Constrained Minimal Supersymmetric Standard Model (CMSSM) is the most widely
studied. 

In this paper, recent results of the LEP experiments, ALEPH, DELPHI, L3 and
OPAL, on the searches for phenomena beyond the SM and
the ``standard" CMSSM are reviewed. These include 
the study of the CMSSM Higgs sector with CP-violation, 
flavour independent searches for Higgs bosons,
Type II Two Higgs Doublet Model (2HDM II) 
interpretation of neutral Higgs boson searches, 
searches for Higgs boson decays to gauge boson pairs,
signatures of Gauge Mediated Supersymmetry Breaking, 
new scalar particles (branons and radions) 
predicted by scenarios with extra spatial dimensions, 
4th generation b' quarks and 
single top quark production.

The results are based on the data collected at LEP2 up to the
highest energies of 209 GeV, corresponding to an integrated luminosity
of around 700 pb$^{-1}$ per experiment.

None of the searches show evidence for new phenomena.  In most cases,
cross-section times branching ratio limits are computed at the 95\% confidence
level (CL) with minimal model assumptions, providing the most general, almost
model independent results. These are then interpreted in the framework of
specific theoretical models to constrain the accessible parameter space and the
properties of the new particles, such as their masses. 

\section{Higgs bosons}\label{sec:Higgs} 

In the SM, the electroweak (EW) symmetry is broken 
via the Higgs mechanism generating the
masses of elementary particles. This requires the introduction of a Higgs field
doublet and implies the existence of a single neutral
scalar particle, the Higgs boson. The minimal extension of the SM Higgs sector,
required for example by supersymmetric models, contains two Higgs field
doublets leading to five Higgs bosons: three neutral and two charged. 

\subsection{CP-violation in the Higgs sector}\label{subsec:CPV} 

In the MSSM, the Higgs potential is assumed to be invariant under CP
transformation at tree level. It is possible, however, to break CP symmetry in
the Higgs sector by radiative corrections. 
Such a scenario could provide a possible
solution to the cosmic baryon asymmetry.

Both CP-conserving (CPC) and CP-violating (CPV) scenarios are studied at LEP. 
In the CPC case, the three neutral Higgs bosons are CP eigenstates: h and H are
CP even, A is CP odd. They are dominantly produced in the Higgs-strahlung
processes  e$^+$e$^-$ $\rightarrow$ hZ, HZ and the pair-production processes
e$^+$e$^-$ $\rightarrow$ hA, HA. In the CPV case, however, the three neutral
Higgs bosons, H$_i$, are mixtures of CP-even and CP-odd Higgs fields and the
processes e$^+$e$^-$ $\rightarrow$ H$_i$Z and  e$^+$e$^-$ $\rightarrow$
H$_i$H$_j$ ($i,j=1,2,3, \ i \ne j$)  may all occur. The decay properties of the
Higgs bosons maintain a certain similarity in the two scenarios: the largest
branching ratios are those to b\=b and $\tau^+\tau^-$, but Higgs-to-Higgs
cascade decays occur and can be dominant when kinematically allowed.

A large number of search channels are used in the MSSM Higgs hunt: SM
Higgs-strahlung  searches are reinterpreted, searches for pair-production 
H$_i$H$_j$, Yukawa production bbH$_i$, flavour independent H$_i$Z and
H$_i$H$_j$ and decay mode independent H$_i$Z searches are developed.
Higgs-to-Higgs decays H$_j$~$\rightarrow$~H$_i$H$_i$ and
H$_j$~$\rightarrow$~H$_i$Z are considered.  The search for invisible decay of
Higgs bosons is used to explore specific parameter regions. In general,
searches designed to detect CPC Higgs production can be reinterpreted in the
CPV scenario. However, modified or newly developed searches are also necessary
to cover new dominating final state topologies, such as 
H$_2$Z~$\rightarrow$~H$_1$H$_1$Z~$\rightarrow$~b\=bb\=bZ with 
$m_{\mathrm{H}_2} \approx 100-110$~GeV.  

The CMSSM has seven parameters. At tree level two parameters are
sufficient to describe the Higgs sector: the ratio of the
vacuum expectation values ($\tan\beta$) and a Higgs mass. 
Additional parameters appear after radiative corrections. 
Instead of varying all the parameters, only a certain number of
representative benchmark sets~\cite{Higgs-benchmarks} are considered. 

In the
CPC benchmark scenarios, the h and A masses are excluded at least up to 87.3
and 93.1 GeV, respectively, except for a tiny region with $\tan\beta < 0.7$ in
the no-mixing scenario. 

\begin{figure}[ht!]
\centering
\includegraphics[width=0.325\textwidth,height=0.3\textwidth,bb=45 5 540 455]{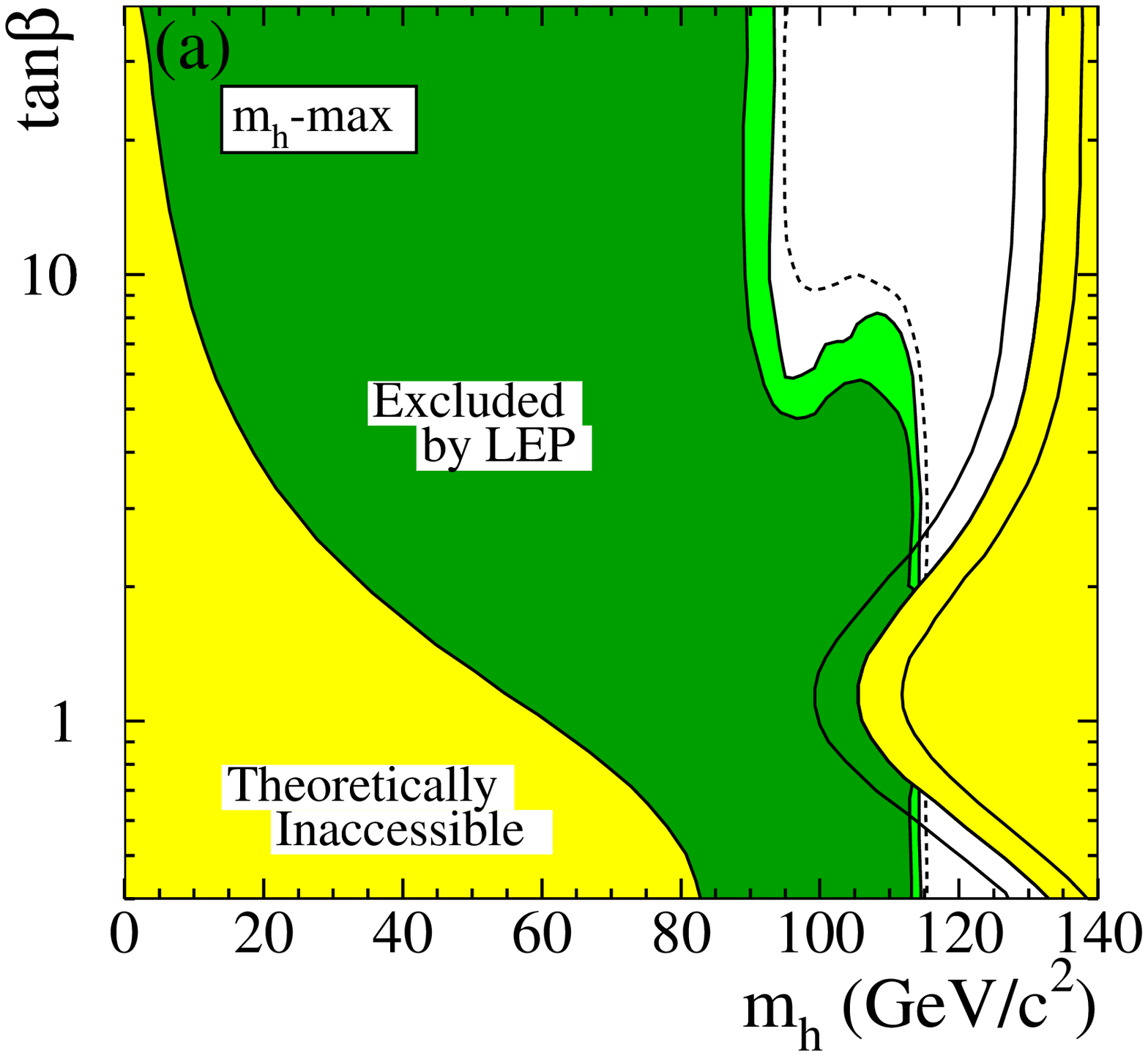}
\hspace*{0.2cm}
\includegraphics[width=0.325\textwidth,height=0.3\textwidth,bb=45 5 540 455]{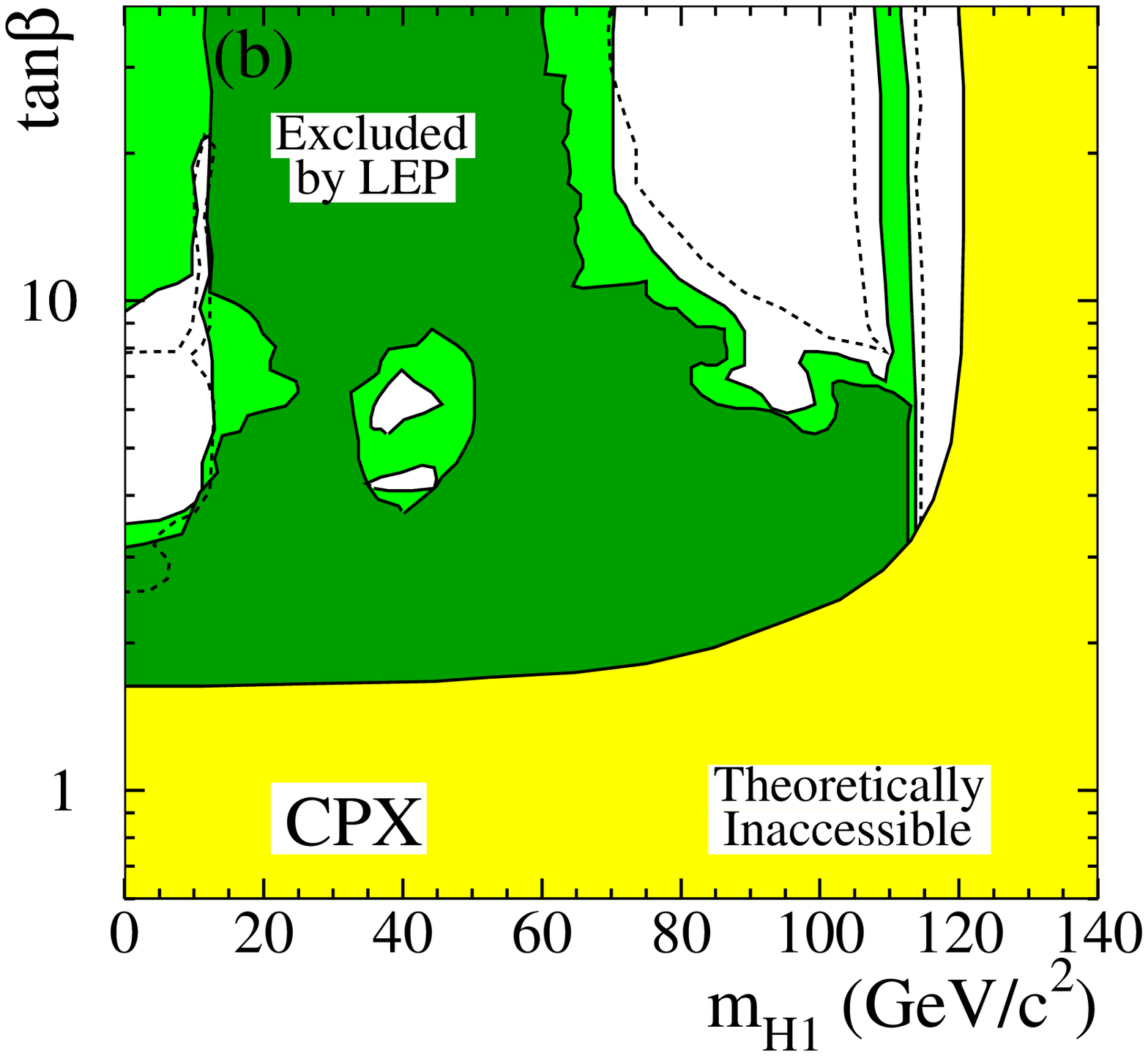}
\vspace*{0.2cm}

\includegraphics[width=0.325\textwidth,height=0.3\textwidth,bb=45 5 540 455]{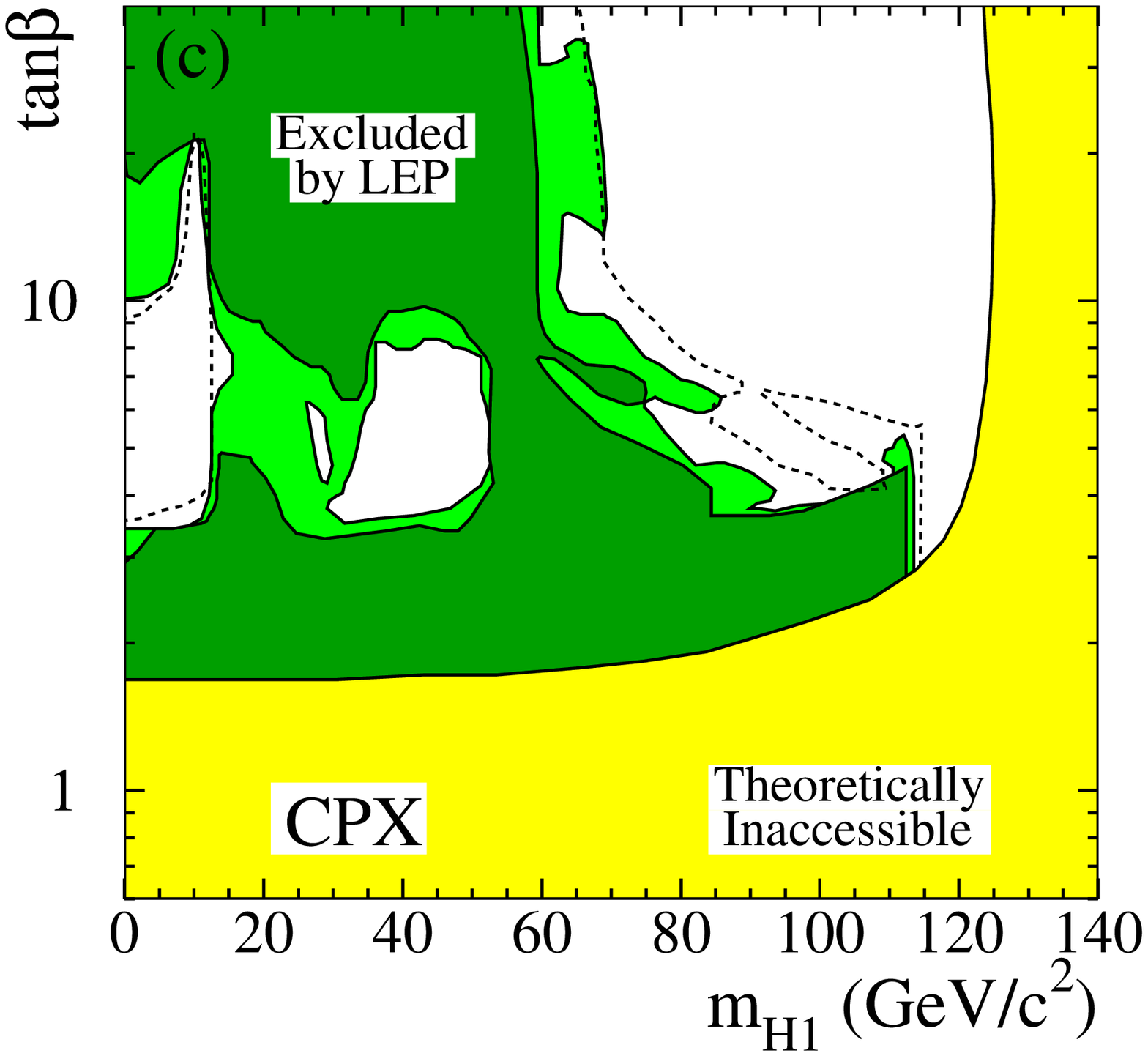}
\hspace*{0.2cm}
\includegraphics[width=0.325\textwidth,height=0.3\textwidth,bb=45 5 540 455]{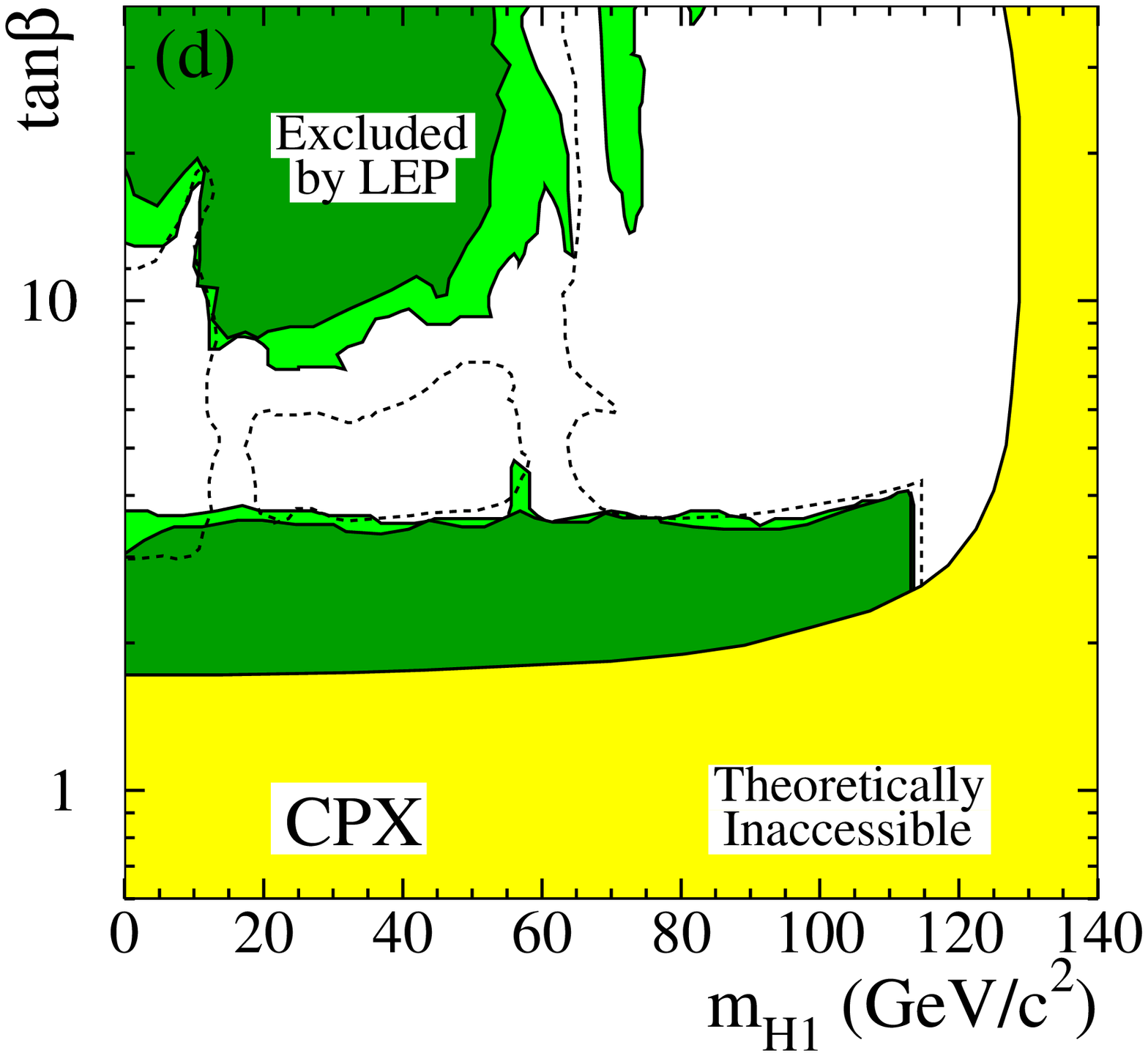}
\caption{Search for the CMSSM neutral Higgs bosons:
Exclusion in the [$m_{\mathrm H_1}- \tan\beta$] plane in the (a) CPC
$m_\mathrm{h}-$max scenario, (b-d) in the CPX scenario for a top quark 
mass of 169.3,
174.3 and 179.3 GeV, respectively. The yellow (light gray) area is theoretically
inaccessible, the dark green (dark gray) region is excluded at the 99\% CL, the
green (gray) region at the 95\% CL, and the dashed line shows the expected
exclusion at the 95\% CL.}
\label{fig:CMSSM}
\end{figure}

The picture dramatically changes, as illustrated on
Figure~\ref{fig:CMSSM}, in the CPV scenario, called CPX~\cite{CPX}, which was
designed to maximize the CPV effects while fulfilling the experimental
constraints from electron and neutron electric dipole moment (EDM)
measurements~\footnote{During the conference the question was raised whether
the recent neutron EDM measurement~\cite{nEDM} help to close the holes in 
the LEP exclusion. 
According to Ref.~\cite{nEDM-Carena}, these results do not constrain further 
the CPX scenario.}. In this scenario H$_1$
decouples from Z in the range $4 < \tan\beta < 10$. H$_2$ couples to the
Z, but it is heavier than around 100 GeV. Where kinematically open, H$_2$
$\rightarrow$ H$_1$H$_1$ is dominant. The excluded areas~\cite{LEP-CMSSM}, by
the combined results of the LEP experiments, are
shown in Figure~\ref{fig:CMSSM}(b-d) for different values of the top quark
mass. It is important to note that while in the CPC case the effect of the top
quark mass is moderate (see the different contours of the theoretically
inaccessible area for large $m_\mathrm{h}$ values on 
Figure~\ref{fig:CMSSM}(a)), the experimental exclusion in the CPV case changes
significantly due to the change in the predicted cross-sections and the ratios
of the Higgs masses with the top quark mass for a given  [$m_{\mathrm H_1}-
\tan\beta$] point.

\subsection{Flavour Independent Searches
and Type II Two Higgs Doublet Models}
\label{subsec:2hdm}

In certain models, for example in the large-$\mu$ CMSSM benchmark or in general
2HDM II, the Higgs coupling to b\=b is suppressed for large regions of the
parameter space. To cover such possibilities, the LEP Collaborations developed
flavour independent selections for H$_i$Z and H$_i$H$_j$ productions, followed
by the decay H$_i$~$\rightarrow$~q\=q, gg. These analyses are experimentally
challenging as it is rather difficult to separate the signal from the
overwhelming WW, ZZ and q\=q(g) backgrounds without the use of the highly
discriminating b- and $\tau$-tagging algorithms. 

\begin{figure}[ht!]
\centering
\includegraphics[width=0.325\textwidth,height=0.3\textwidth,bb=10 25 430 420]{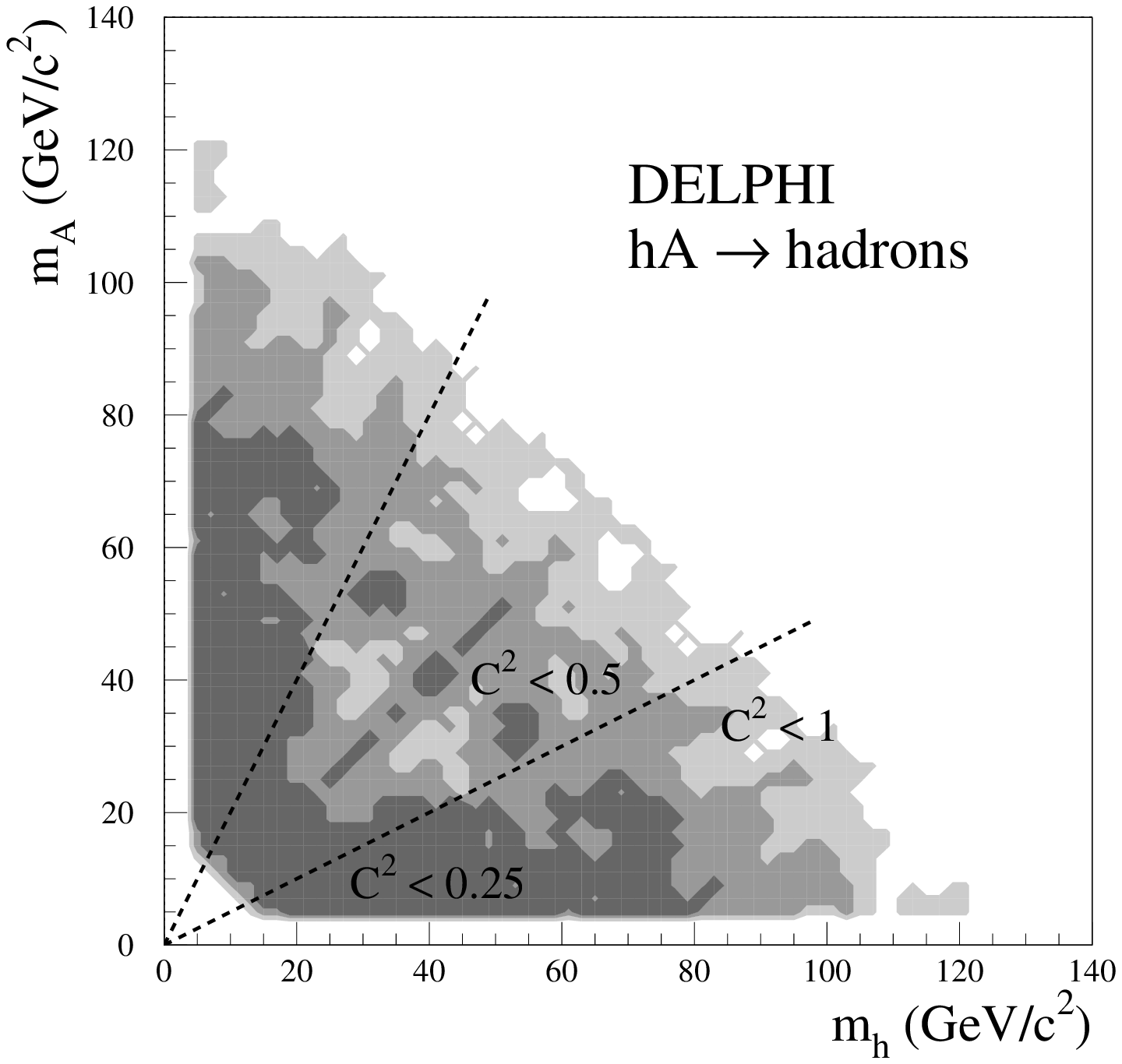}
\hspace*{0.4cm}
\includegraphics[width=0.325\textwidth,height=0.3\textwidth,bb=0 0 515 490]{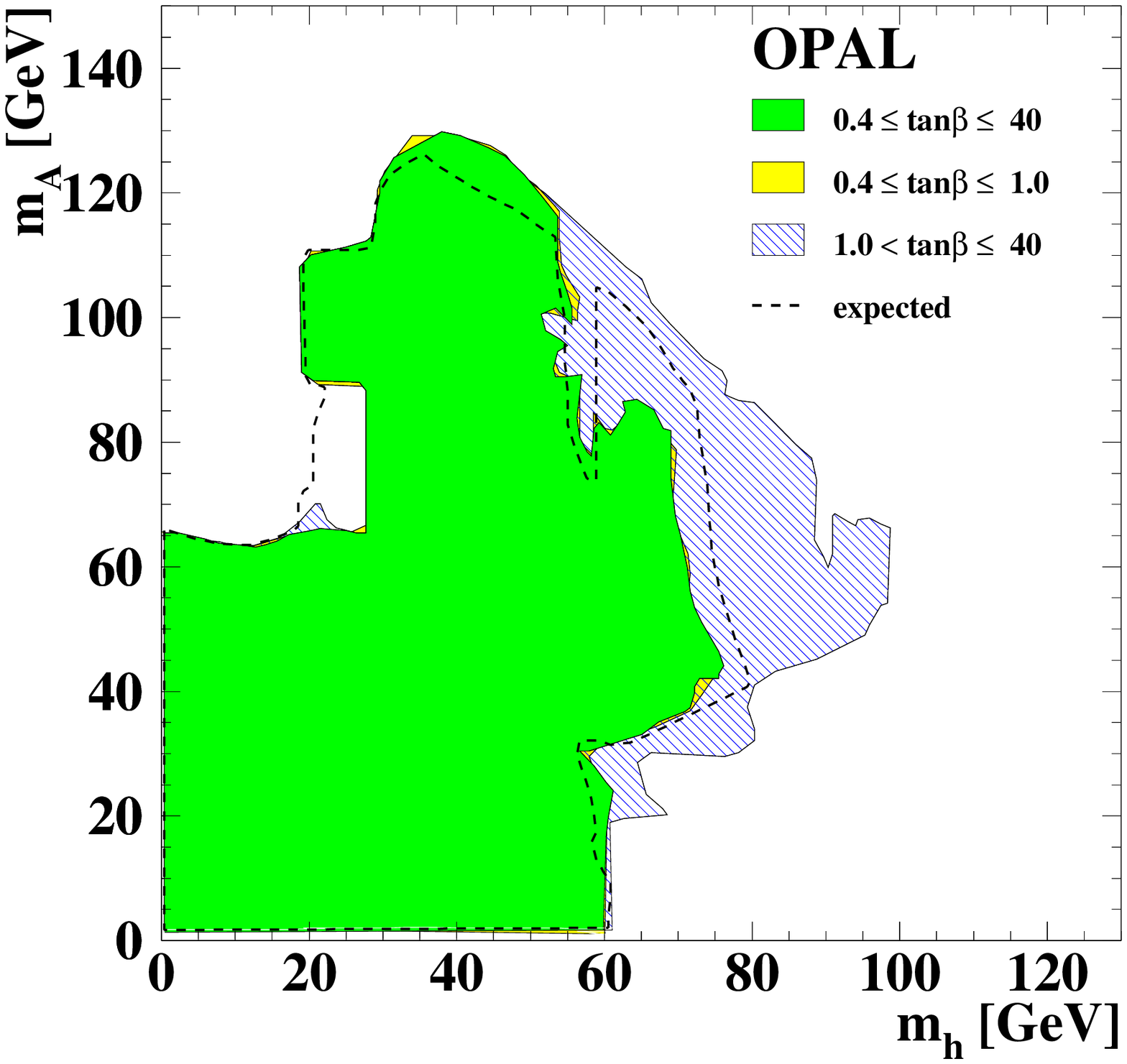}
\caption{
(a) Flavour independent search for the pair-production of neutral Higgs bosons: 
Upper bounds on the
parameter $C^2$ in the [$m_{\mathrm h} - m_{\mathrm A}$] 
(or equivalently [$m_{\mathrm H_2} - m_{\mathrm H_1}$]) plane. 
(b) Interpretation of neutral Higgs boson searches in 2HDM II: Excluded areas in
the [$m_{\mathrm h} - m_{\mathrm A}$] plane, independent of $\alpha$.}
\label{fig:2HDM}
\end{figure}

The pair-production process
was sought by the DELPHI~\cite{DELPHI-flavindep} and OPAL~\cite{OPAL-2HDM}
Collaborations.  The results are expressed as limits on the cross-section
scaling parameter $C^2$, which is defined to be 1 for the
maximal production cross-section allowed by EW symmetry breaking and for 100\%
decay into hadrons. Figure~\ref{fig:2HDM}(a) shows such a model-independent
presentation from DELPHI.

The OPAL Collaboration interpreted the results of neutral Higgs boson searches
in the framework of a general 2HDM II model, assuming CP-conservation in the
Higgs sector and no additional non-SM particles other than the Higgs
bosons~\cite{OPAL-2HDM}. The parameters $m_{\mathrm h}$, $m_{\mathrm A}$,
$\tan\beta$ and $\alpha$ (the mixing angle in the neutral CP-even Higgs sector)
were scanned and the other two free parameters $m_{\mathrm H}$ and $m_{\mathrm
H^\pm}$ were set above the kinematically accessible region. The excluded
areas are shown in  Figure~\ref{fig:2HDM}(b).

\subsection{Fermiophobic Higgs bosons}\label{subsec:HtoWW} 

In the SM the Higgs branching ratio to heavy gauge bosons increases
with mass, but even at the LEP2 kinematic limit, it accounts for less than 10\%
of the decays. In certain extensions (for example in Type I Two Higgs Doublet
Models), the Higgs boson may become fermiophobic and its decay to gauge bosons
may become dominant. 

\begin{figure}[ht!]
\centering
\includegraphics[width=0.325\textwidth,height=0.3\textwidth,bb=0 10 500 565]{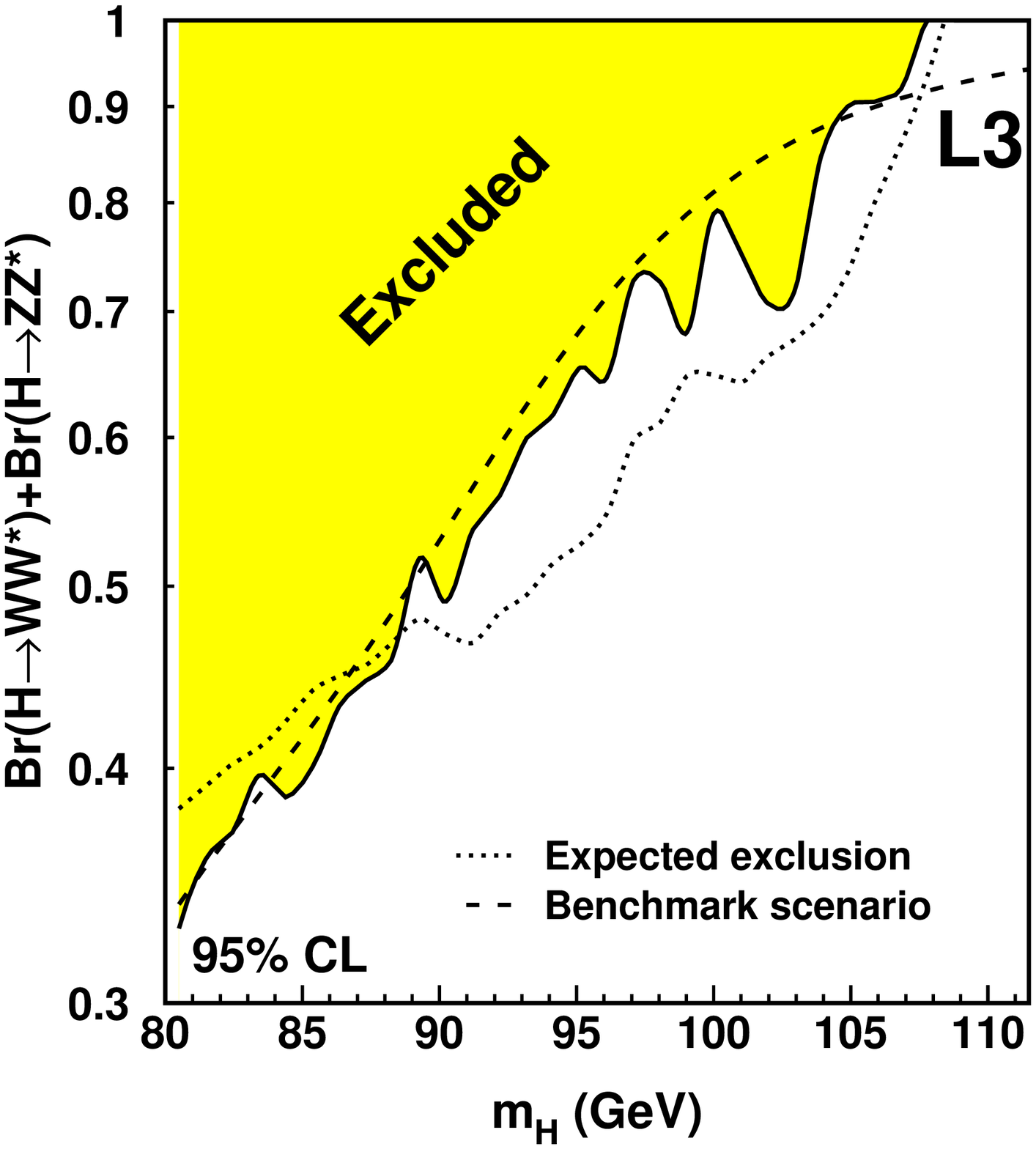}
\hspace*{0.4cm}
\includegraphics[width=0.325\textwidth,height=0.3\textwidth, bb=20 5 555 500]{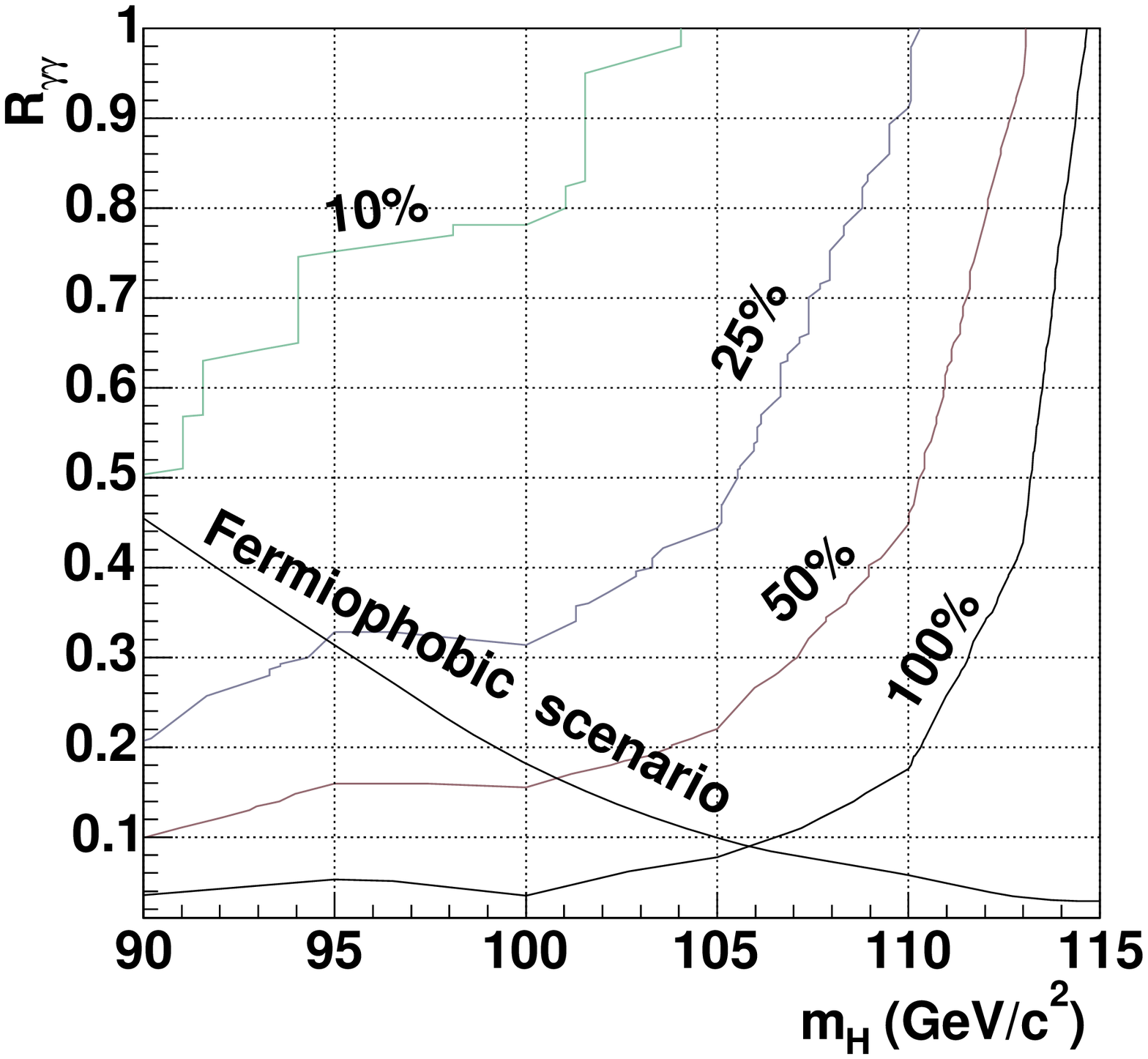}

\vspace*{-4.6cm}
\hspace*{3.8cm} ALEPH
\vspace*{4cm}

\caption{Search for fermiophobic Higgs bosons:
(a) Excluded region in the [$m_{\mathrm H} - Br$(H$\rightarrow$WW*, ZZ*)] plane.
(b) Upper bounds on the total decay rate to electroweak gauge bosons 
($Br_{\mathrm boson}$) as a function of $m_{\mathrm H}$ and  the ratio of the
H$\rightarrow \gamma\gamma$ decay rate to the total bosonic decay rate
($ R_{\gamma\gamma} = Br_{\gamma\gamma} / Br_{\mathrm boson}$).} 
\label{fig:HtoWW}
\end{figure}

The ALEPH and L3 Collaborations performed searches for H $\rightarrow$ WW* (ZZ*) 
decays~\cite{HtoWW}. The resulting numerous six fermion final states are
grouped into different topological classes depending on the number of hard and
soft leptons, jets and the amount of missing energy. The L3
results are given on Figure~\ref{fig:HtoWW}(a). These can be combined with the
searches for H $\rightarrow \gamma\gamma$, see the example from ALEPH on
Figure~\ref{fig:HtoWW}(b), to improve the sensitivity to the fermiophobic
benchmark scenario, which assumes SM-like Higgs couplings to bosons and no
coupling to fermions. 

\section{Signatures of Gauge Mediated Supersymmetry Breaking}\label{sec:gmsb}

Supersymmetry (SUSY), the best proposed solution to the problems of the
SM, postulates the existence of a partner for each SM particle chirality
state. The discovery of these superpartners would be the most direct evidence
for SUSY. Since SUSY particles are not observed with the same mass as their SM
partners, SUSY must be broken. In the most widely investigated scenarios, it is
assumed that SUSY is broken in some {\em hidden} sector of new particles and is
{\em communicated} to the {\em visible} sector of SM and SUSY particles by
gravity or gauge interactions. 

We review here a study of gauge-mediated SUSY breaking (GMSB)
topologies~\cite{OPAL_GMSB} using the data collected by the OPAL detector.

In models with GMSB, the lightest SUSY particle (LSP) is a light gravitino
$\tilde\mathrm{G}$. The
phenomenology is driven by the nature of the next-to-LSP (NLSP), which is
either the  lightest neutralino $\tilde\chi^0_1$, scalar tau $\tilde\tau_1$ or 
mass-degenerate scalar leptons $\tilde\ell$. As the gravitino couples very
weakly to heavier SUSY particles, those will decay typically to the NLSP which
then decays via  $\tilde\chi^0_1 \rightarrow \gamma \tilde\mathrm{G}$ or 
$\tilde\ell \rightarrow \ell \tilde\mathrm{G}$. All relevant final states are
considered: direct NLSP production and  its appearance in the decay chain of
heavier SUSY particles, like charginos, neutralinos and scalar leptons.

The minimal GMSB model introduces five new parameters and a sign:  the SUSY
breaking scale ($\sqrt{F}$), the SUSY particle mass scale ($\Lambda$), the
messenger mass ($M$), the number of messenger sets ($N$), $\tan\beta$ and
the sign of the SUSY Higgs mass parameter (sign$(\mu)$). 
As the decay length of the NLSP depends on $\sqrt{F}$ and is
effectively unconstrained, NLSP decays inside and outside of the detector are
searched for. With increasing decay length, the event signatures include: 
energetic leptons or photons and  missing energy due to the undetected
gravitino, tracks with large impact parameters, kinked tracks, or heavy
long-lived charged particles. In total more than 14 different selections are
developed to cover the GMSB topologies. The results are
combined (with special attention to treat the overlaps among the many channels
properly) to get lifetime independent results, eliminating the dependence on 
$\sqrt{F}$. 

None of the searches shows evidence for SUSY particle production. To interpret
the results, a detailed scan of the minimal GMSB parameter space is performed. 

``Model independent" cross-section limits are derived for each topology as a
function of the NLSP lifetime, both for direct NLSP production and, for the
first time, also for cascade decays.  For direct NLSP production, 
this is done by 
taking the worst limit for a given NLSP mass from the generated GMSB parameter
scan points. For cascade channels, the cross-section evolution is  assumed to
be $\beta/s$ for spin-1/2 and $\beta^3/s$  for scalar SUSY particles,
and the highest bound for all intermediate particle masses is
retained. The maximum limit valid for all lifetimes is then quoted as the
``lifetime independent" cross-section limit. In the neutralino NLSP scenario
this is typically better than 0.04 pb for direct NLSP production, 0.1 pb for
scalar electron and scalar muon production, 0.2 pb for scalar tau production
and 0.3 pb for chargino production. In the scalar tau and scalar lepton co-NLSP
scenarios, the limit on direct NLSP production cross-section is smaller than
about 0.05 pb for scalar muons, 0.1 pb for scalar electrons and scalar taus.
For the cascade decays the bounds are typically better than 0.1 pb for
neutralino, 0.2 for chargino and, in the scalar tau NLSP scenario, 
0.4 for scalar electron and scalar muon production.

\begin{figure}
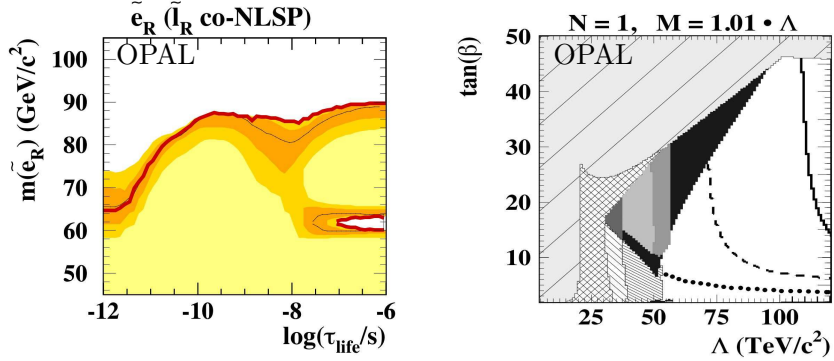

\centering
\includegraphics[width=0.325\textwidth, height=0.3\textwidth,bb=0 0 270 250]{pr409_19.epsi.epsf}
\hspace*{0.4cm}
\includegraphics[width=0.325\textwidth, height=0.29\textwidth,bb=0 5 220 230]{pr409_20.epsi.epsf}

\vspace*{-4.4cm}
OPAL \hspace*{4.5cm} OPAL \hspace*{1cm}
\vspace*{3.7cm}

\caption{ Search for GMSB signatures:
(a) Observed (thick red/dark gray) and  expected (thin black) lower mass limits
for pair-produced scalar electrons in the scalar lepton co-NLSP scenario as a
function of the NLSP lifetime. The
68\% and 95\% probability intervals are shown by orange/grey shades. 
(b) Regions in the [$\Lambda - \tan\beta$] plane excluded by the different
searches in the scalar tau and scalar lepton co-NLSP scenarios (direct NLSP
production - black, chargino - dark gray, neutralino - gray, scalar electron
and scalar muon production with scalar tau NLSP - light gray) and in the
neutralino NLSP scenario (neutralino production - dense hatched, chargino -
hatched). The  LEP1 search region is cross-hatched and the theoretically not
allowed area is sparse hatched gray. The constraint from the LEP
combined Higgs limit of $m_\mathrm{H}>114.4$~GeV is also shown (full line)
together with the indication of the large effect from theoretical (dashed)
plus top quark mass (dotted)  uncertainties, weakening this constraint.}
\label{fig:gmsb}
\end{figure}

The cross-section limits can be turned into constraints on the NLSP
mass. For scalar leptons, the lowest mass limits are found for very short
lifetimes, except for scalar electrons, shown in Figure~\ref{fig:gmsb}(a), where 
searches using d$E$/d$x$ measurements loose efficiency for particles with
momenta around 65 GeV. The lifetime independent limits are 
$m_{\tilde\mathrm{e}_\mathrm{R}} > 60.1$~GeV, $m_{\tilde\mu_\mathrm{R}} >
93.7$~GeV and  $m_{\tilde\tau_1} > 87.4$~GeV. 
The limit on $m_{\tilde\tau_1}$  is the
same in the scalar tau and the scalar lepton co-NLSP scenarios. 
In the scalar lepton co-NLSP
scenario, where by definition the mass differences between the different
scalar lepton flavors are smaller than the lepton masses,
the best limit can be used to derive a universal limit on 
$m_{\tilde\ell} = m_{\tilde\mu_\mathrm{R}} - m_\tau > 91.9$~GeV. 
For neutralino NLSP, no
lifetime independent NLSP mass limit can be set directly. For short lifetimes
($\tau < 10^{-9}$~s), a mass limit of $m_{\tilde\chi_1^0} > 96.8$~GeV 
is derived.

The GMSB parameter space can also be constrained by these results as illustrated
in
Figure~\ref{fig:gmsb}(b) for $N=1$, $M=1.01\cdot\Lambda$ and sign$(\mu) > 0$.
The universal SUSY mass scale is $\Lambda> 40,\, 27,\, 21,\, 17,\, 15$~TeV for 
messenger indices $N=1,\, 2,\, 3,\, 4,\, 5$, independent of the other model
parameters. The constraints on $\Lambda$
imply a lifetime independent lower limit on the neutralino mass in the
neutralino NLSP scenario: $m_{\tilde\chi_1^0} > 53.5 \ (94.0)$~GeV for 
$N = 1 \ (5)$.

\section{Extra Dimensions}\label{sec:ed}

Models with extra spatial dimensions (EDs) have been introduced to solve the
hierarchy problem of the SM through geometrical considerations.
Most LEP results are derived in the original Arkani-Hamed -- 
Dimopoulos -- Dvali (ADD) framework, which assumes $n$ compact
EDs, with the Planck scale $M_D$ in $D=4+n$ dimensions set close
to the EW scale. SM particles propagate in a four dimensional (4D) 
subspace (brane),
while gravity in the full $D$ dimensional space. The 4D Planck
scale $M_{\mathrm  Planck}$ satisfies $M_{\mathrm  Planck}^2 = V_n
M_D^{n+2}$, where $V_n$ is the volume of the EDs.

An other interesting scenario was suggested by Randall and Sundrum. They assume
only one ED and generate the hierarchy by a specifically chosen
``warped" geometry. Gravity is then located close to a second brane and its
propagation in the ED is exponentially damped.  

A general prediction of these scenarios is the existence of massive 
Kaluza-Klein
excitations of the graviton in the 4D effective theory. 

\subsection{Branons}\label{subsec:branons}

In a general ADD like geometry, where SM particles live on a 3-brane, the
presence of brane fluctuations of a typical size $1/f$ manifests themselves as
new scalar weakly interacting particles, the branons, which also serve as
possible dark matter candidates. If the branes are flexible, i.e. the brane
tension $f$ is much smaller than the D dimensional Planck scale, the graviton
KK modes decouple from SM particles and the first signal from the EDs could be
the discovery of branons. 

\begin{figure}[ht!]
\centering
\includegraphics[width=0.3\textwidth,height=0.3\textwidth,bb=5 -10 515 515]{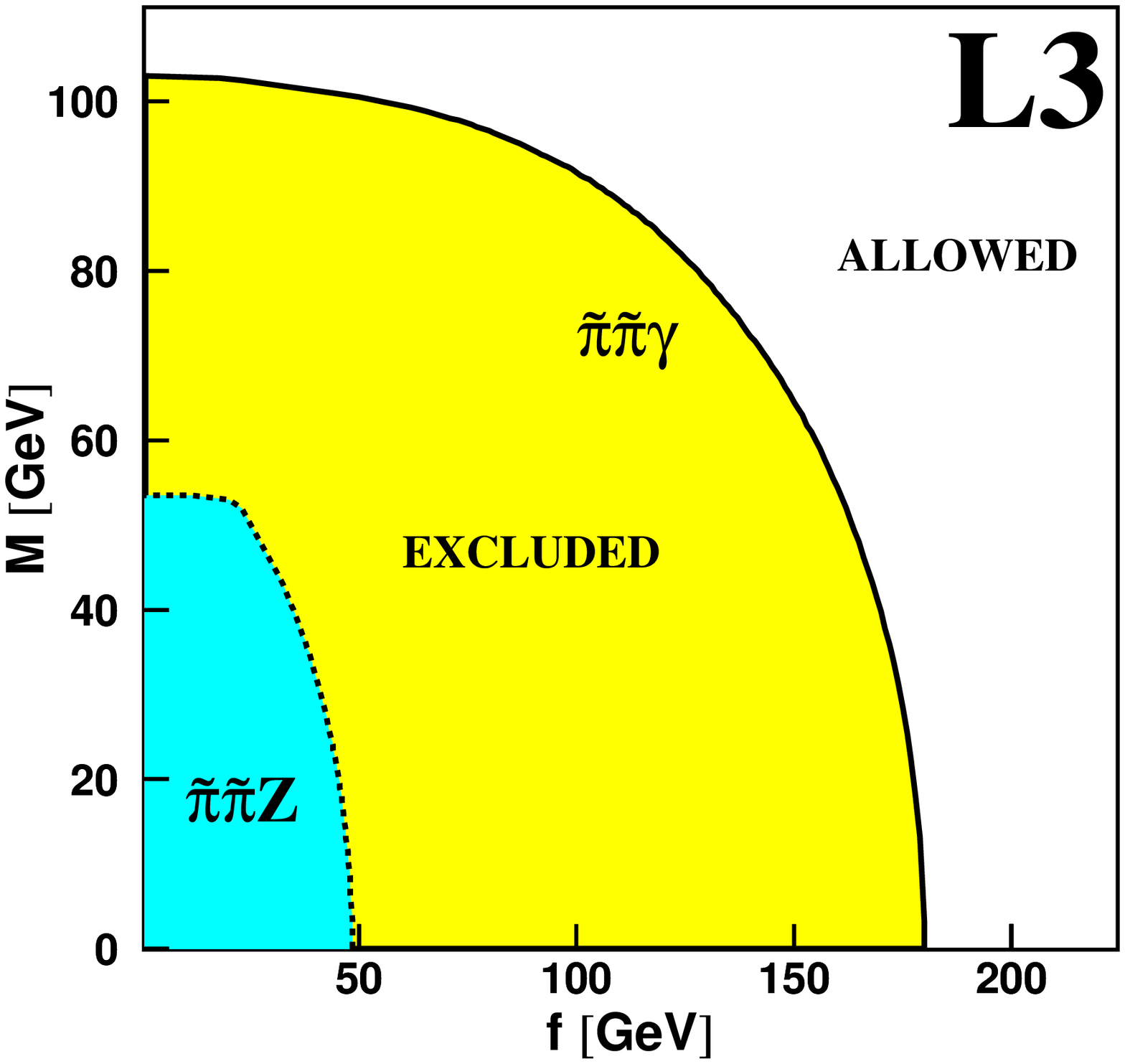}
\hfill
\includegraphics*[width=0.36\textwidth,height=0.33\textwidth,bb=295 13 590 233]{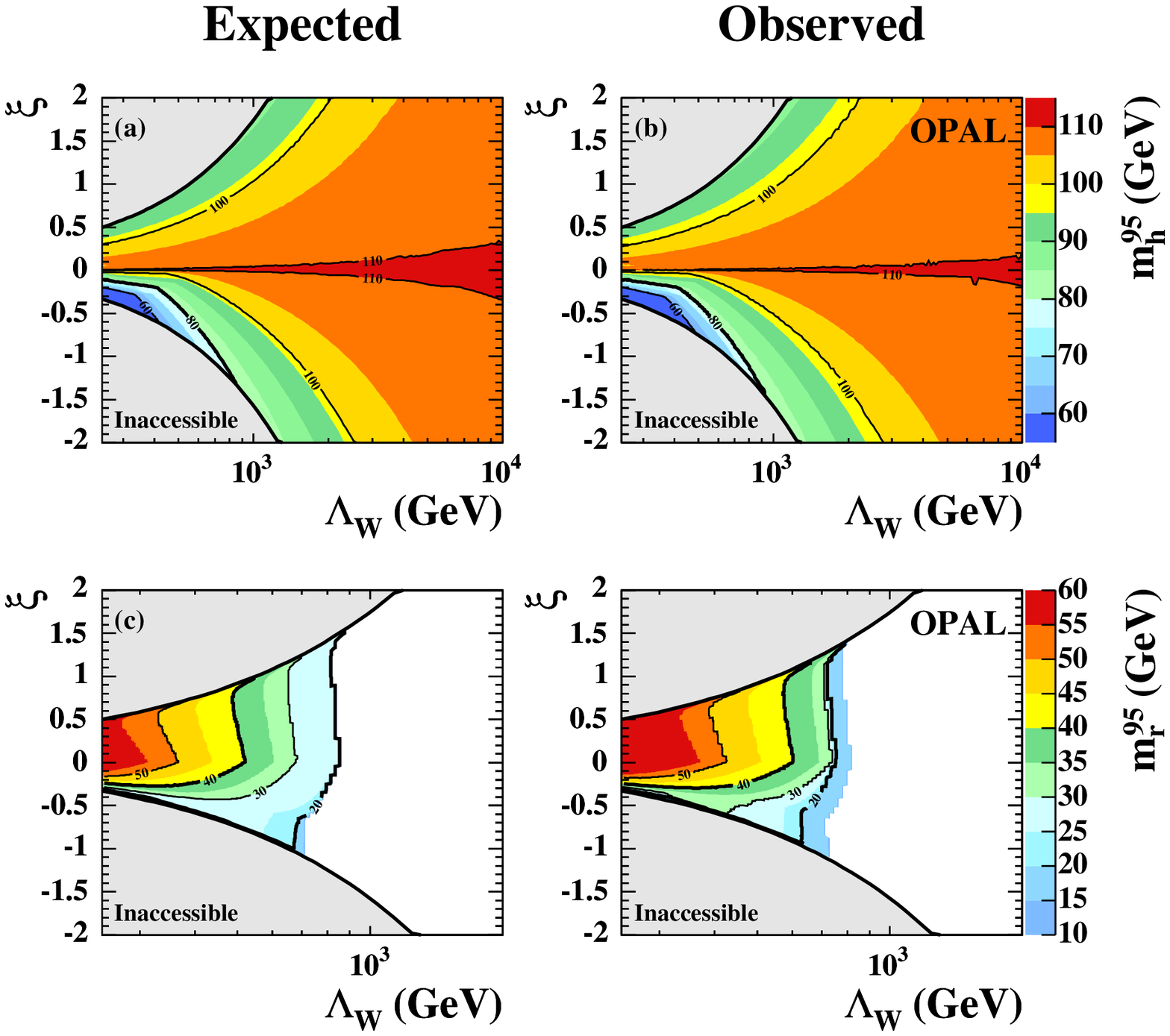}
\hfill
\includegraphics[width=0.3\textwidth,height=0.3\textwidth,bb=0 -8 232 230]{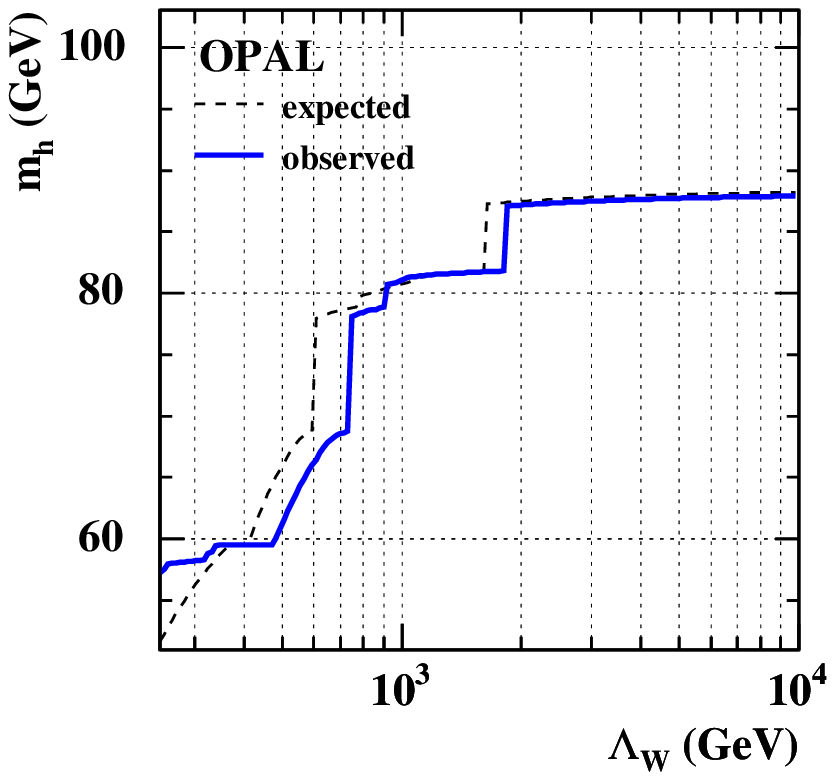}
\caption{(a) Search for branons: Excluded region in the [brane tension - branon
mass] plane, assuming only one branon of mass M.
Search for radions: (b) Lower limit on the mass of the radion-like state as a
function of $\Lambda_\mathrm{W}$ and $\xi$. (c) Absolute lower limit on the
mass of the Higgs-like state as a function of $\Lambda_\mathrm{W}$.}
\label{fig:ED}
\end{figure}

Branons ($\tilde\pi$) couple in pairs to SM particles,
and they would appear in e$^+$e$^-$ collisions in Z/$\gamma$ $\tilde\pi
\tilde\pi$ final states, giving a single Z or $\gamma$ plus missing energy. 
The limits on branon production from the L3 Collaboration~\cite{L3-branons} are
shown on Figure~\ref{fig:ED}(a).

\subsection{Radions in the Randall-Sundrum Model}\label{subsec:radions}

In the Randall-Sundrum (RS) scenario, the radion corresponds to local
fluctuations of the inter-brane distance. The radion has the same quantum
numbers as the Higgs boson and mixes with it, resulting in a radion-like (r)
and a Higgs-like (h) state. The radion couplings are similar to the Higgs
couplings but suppressed by a factor of $v/(\sqrt{6}\Lambda_\mathrm{W})$, where
$v$ is the vacuum expectation value of the Higgs field and $\Lambda_\mathrm{W}$
is the energy scale on the SM brane. The radion, however, also couples to gluon
pairs, and the r $\rightarrow$ gg decay is dominant. 

The OPAL Collaboration re-interpreted its SM, flavour and decay-mode
independent Higgs boson searches in the RS model and derived limits on the r
and h masses as a function of $\Lambda_\mathrm{W}$ and the mixing parameter,
$\xi$, as shown for the radion-like state on Figure~\ref{fig:ED}(b). As opposed
to the Higgs-like state, searches for the radion-like state loose
sensitivity for larger values of  $\Lambda_\mathrm{W}$ and for large negative
values of $\xi$ close to the theoretically inaccessible region, therefore, no
absolute limit on the mass can be derived. The absolute lower limit on the mass
of the Higgs-like state is given on Figure~\ref{fig:ED}(c).

\section{Fourth Generation b' Quarks}\label{sec:b'}

In the SM, the number of fermion generations and their mass spectrum are not
predicted. Precision EW measurements allow for the existence of an extra, heavy
fermion generation, if $|m_\mathrm{t'} - m_\mathrm{b'}| < 60$~GeV is fulfilled.
For $m_\mathrm{Z} < m_\mathrm{b'} < m_\mathrm{H}$, the b' quark decays
predominantly into bZ and cW. 

\begin{figure}[hbt!]
\centering
\includegraphics[width=0.325\textwidth,height=0.3\textwidth,bb=15 15 298 235]{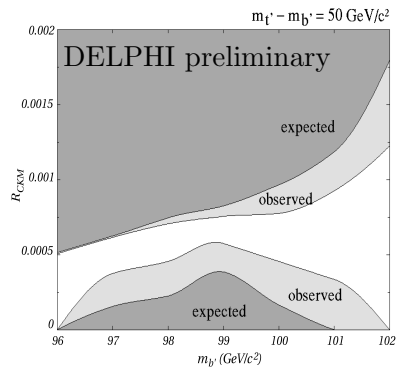}

\vspace*{-4.5cm}
DELPHI preliminary
\vspace*{3.9cm}

\caption{Search for b' quarks: Excluded regions in the 
[$m_{b'} - R_\mathrm{CKM}$] plane for $m_\mathrm{t'} - m_\mathrm{b'} = 50$~GeV . 
The constraints on $R_\mathrm{CKM} = | V_{cb'} / ( V_{tb'} V_{tb} ) |$ weaken 
as the mass difference $m_\mathrm{t'} - m_\mathrm{b'}$ decreases.}
\label{fig:bp}
\end{figure}

The DELPHI Collaboration performed a search for b'b' production in
the bZbZ and cWcW final states for $m_\mathrm{b'} = 96 - 103$~GeV, and
derived upper limits on the b' $\rightarrow$ bZ and 
b' $\rightarrow$ cW branching ratios of 51
and 43\% for $m_\mathrm{b'}=96$~GeV, degrading to 74 and 55\% for
$m_\mathrm{b'}=101$~GeV,  respectively~\cite{DELPHI-bp}. The branching ratio
limits can be used to constrain the extended CKM matrix within a
four-generation sequential model, as shown on Figure~\ref{fig:bp}.

\section{Single Top Quark Production via Contact Interactions}\label{sec:singletop}

As flavour changing neutral currents (FCNCs) are forbidden at tree level in the
SM, in good agreement with the experimental data, rare FCNC processes are ideal
to look for new physics. The LEP Collaborations conducted searches 
for single top quark production via FCNC, described in terms of vector-like anomalous
couplings ($\kappa_\gamma, \kappa_\mathrm{Z}$) associated with the photon and
the Z
boson.

A more general approach is to consider possible new 4-fermion contact
interactions, including scalar-, vector- and tensor-like couplings, through an
effective Lagrangian with FCNC operators. The eetc vertex is 
then characterized by
the couplings $S_\mathrm{RR}, V_{ij}$ and $T_\mathrm{RR}$, where $i,j =$ L,R.
To account for the 'traditional' FCNC Ztc vertex, the
couplings $a_j^\mathrm{Z}$ are also introduced.

The DELPHI Collaboration considered several scenarios in this framework defined
by the couplings that are different from zero: S, V, T, SVT, a, V$-$a, V+a, where
the $-/+$ signs refer to destructive or constructive interference between the
vector like 4-fermion and the Ztc couplings. From the search for the 
e$^+$e$^-$ $\rightarrow$ t\=c $\rightarrow$ b\=cq\=q' and  b\=c$\ell\nu_\ell$
processes, they placed bounds on the physics scale $\Lambda >$ 0.69,
1.07, 1.20, 1.40, 0.50, 1.09, 1.06 TeV, in the six scenarios listed
above~\cite{DELPHI-singletop}. These results are consistent with earlier
results of the L3 Collaboration in the S, V, T scenarios~\cite{L3-singletop}.
When only one coupling is different from zero, it can also be constrained from
the data: $S/\Lambda^2 < 2.14 \cdot 10^{-6}$ and 
$T/\Lambda^2 < 6.90 \cdot 10^{-7}$ GeV$^{-2}$.

\section{Conclusion}\label{sec:summary}

The LEP machine was an ideal tool to search for physics beyond the SM, and a
huge number of such scenarios were studied and constrained. The LEP results,
both on precision EW measurements and direct searches, give us hints how to
continue the quest to uncover a more fundamental theory of particle physics at
the LHC and beyond.



\end{document}